\documentclass[a4paper, 11pt]{article}

\pdfoutput=1

\usepackage{graphicx}

\def\b{\begin{eqnarray}}
\def\e{\end{eqnarray}}
\def\n{\noindent}
\begin{document}

\begin{center}

{\huge \textbf{Daemon Decay and Inflation}}

\vspace {10mm}
\noindent
{\large \bf Emil M. Prodanov} \vskip.5cm
{\it School of Mathematical Sciences, Dublin Institute of Technology, Ireland,} \vskip.1cm
{\it E-Mail: emil.prodanov@dit.ie} \\
\vskip1cm
\end{center}

\vskip4cm
\begin{abstract}
\n
Quantum tunneling in Reissner--Nordstr\"om geometry is studied and the tunneling rate is determined.
A possible scenario for cosmic inflation, followed by reheating phases and subsequent radiation-domination expansion, is proposed.
\end{abstract}

\newpage
\n
In 1971, Hawking suggested \cite{hawk} that there may be a very large number of gravitationally collapsed charged objects of very low masses, formed as a result of fluctuations in the early Universe. A mass of $10^{14}$ kg of these objects could be accumulated at the centre of a star like the Sun. The masses of these collapsed objects are from $10^{-8}$ kg and above and their charges are up to $\pm$ 30 electron units \cite{hawk}. \\
Tracing the evolution of such objects, we propose a mechanism that accounts for the cosmic
inflation, takes us into a period of reheating phases, and, finally, into the expansion of a radiation-dominated Universe. In a nut-shell, the inflation mechanism is based on the accumulative effects of Coulomb repulsion at very short range, initially completely ``cocooned'' by Reissner--Nordstr\"om gravitational effects and subsequently unleashed by quantum tunneling. \\
Consider the Reissner--Nordstr\"om geometry \cite{rn, mtw} in Boyer--Lindquist coordinates \cite{bl}:
\b
ds^2 & = & - \, \, \frac{\Delta}{r^2} \, dt^2 + \frac{r^2}{\Delta} \, dr^2  +
r^2 \, d \theta^2 + r^2 \sin^2 \!\theta \, d\phi^2 \, .
\e
where: $\Delta  =  r^2 - 2 M r + Q^2 \, , \, $  $M$ is the mass of the centre, and $Q$ --- the charge of the centre.
We will be interested in the case of a naked singularity only, namely: $\vert Q \vert > M$. \\
The radial motion of an ultra-relativistic test particle of mass $m$ and charge $q$ in Reissner--Nordstr\"om geometry can be modeled by an effective one-dimensional motion of a particle in non-relativistic mechanics with the following equation of motion \cite{pig1, pig2, cohen} (see also \cite{wald} for Schwarzschild geometry) :
\b
\label{w}
\frac{\dot{r}^2}{2} +  U(r) = \frac{\epsilon^2 - 1}{2} \, ,
\e
where
\b
\label{pot}
U(r) = \frac{1}{2}\Bigl( 1 - \frac{q^2}{m^2} \Bigr) \frac{Q^2}{r^2}
- \Bigl( 1 - \frac{q}{m} \, \frac{Q}{M} \, \epsilon \Bigr) \frac{M}{r}
\equiv - \frac{a}{r^2} + \frac{b}{r}
\e
is the effective non-relativistic one-dimensional potential per unit mass, $E = (\epsilon^2 -1)/2$ is the specific energy of the effective one-dimensional motion, and $\epsilon = kT/m + 1 \, $ is the specific energy of the three-dimensional relativistic motion. In equation (\ref{pot}), the constant $a = - Q^2( 1 - q^2/m^2)/2 \, $ is positive in view of the very high charge-to-mass ratio $q/m$ for all charged elementary particles and the parameter $b = - M [1 -(qQ\epsilon)/(mM)]$ depends on the temperature via $\epsilon$. Motion is allowed only when the kinetic energy is real. Equation (\ref{w}) determines the region $(r_- \, , r_+)$ within which classical motion is impossible. The turning radii are given by \cite{pig1, pig2, cohen}:
\b
\label{r0}
r_\pm = \frac{M}{\epsilon^2 - 1} \Biggl[ \epsilon \, \frac{q}{m} \, \frac{Q}{M} - 1
\pm \sqrt{\Bigl( \epsilon \, \frac{q}{m} \, \frac{Q}{M} - 1 \Bigr)^2
- (1 - \epsilon^2) \Bigl(1 - \frac{q^2}{m^2} \Bigr) \frac{Q^2}{M^2}} \, \, \Biggr].
\e
There is no inner turning radius $r_-$ for particles of specific charge $q/m$ such that
sign$(Q)q/m < 1$. For particles such that sign$(Q)q/m < -1$, there is neither inner turning radius, nor outer turning radius \cite{pig1, pig2, cohen}. Such particles will fly unopposed into the centre. Barrier with {\it two turning radii} is present only for particles for which sign$(Q)q/m \ge 1$. We will consider only such particles. Thus the parameter $b$ will be taken as positive.
\begin{center}
\includegraphics[width=8cm]{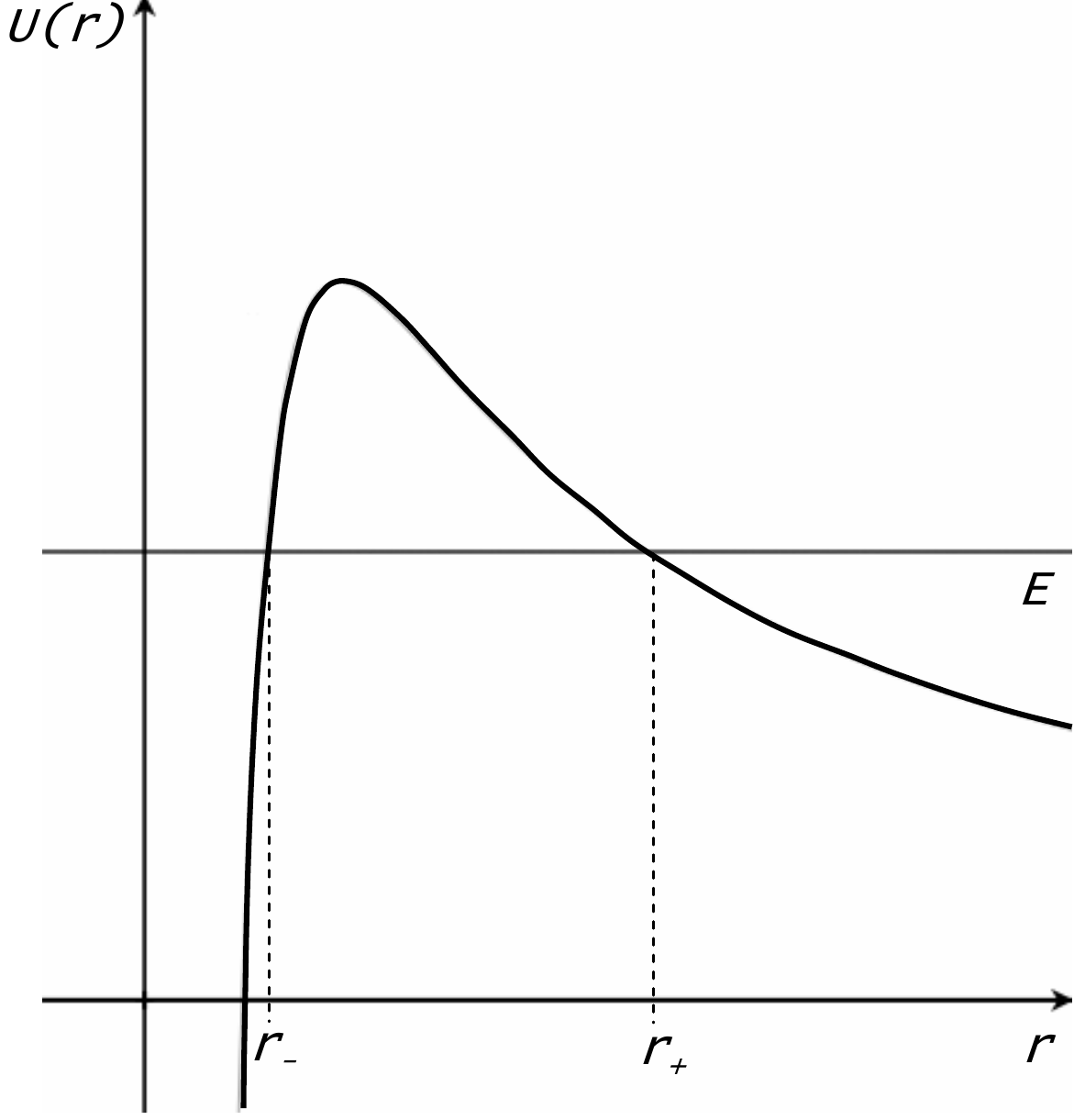}
\vskip.6cm
\noindent
\parbox{110mm}{\begin{center}\footnotesize  Figure 1: The effective potential
\vskip.3cm
$U(r) = \frac{1}{2}\Bigl( 1 - \frac{q^2}{m^2} \Bigr) \frac{Q^2}{r^2}- \Bigl( 1 - \frac{q}{m} \, \frac{Q}{M} \, \epsilon \Bigr) \frac{M}{r} \equiv - \frac{a}{r^2} + \frac{b}{r}$
\end{center}}
\end{center}
There is no classical analogue of this effect: a charged centre being able to capture particles of the same charge within the inner turning radius $r_-$, despite of the Coulomb repulsion. We make the following assumption: the pre-inflationary Universe is an ideal quantum gas in thermal equilibrium with constant volume densities of the positive and the negative charges. Under minute density fluctuation in the volume density of one type of charges at some point, the domain of all other like charges within radius $r_-$ are trapped gravitationally into a cluster. We will call this domain a daemon (for dark electric matter objects, as introduced by \cite{drob}, and in line with our previous work \cite{pig1, pig2} on Reissner--Nordstr\"om expansion). There will be no charges of this type between $r_-$ and $r_+$, while charges of this type on the outside of $r_+$ would be strongly repelled. As a result, the pre-inflationary Universe nucleates into such domains (daemons). Domains of different charge can, obviously, overlap: an oppositely charged particle, approaching a daemon, will not experience turning radii, will fly into the daemon and freely interact with the particles in it. As our aim is to give a qualitative description, we also assume that $m$ is the typical mass of an elementary particle, while $q$ is its typical charge. \\
It should be noted that when both turning radii are present, they are always real, that is, that they are real for any value of $\epsilon$ (or any temperature). The discriminant $(4Q^2/M^2)(1 - q^2/m^2)(Q^2/M^2-1)$ of the quadratic expression in $\epsilon$ under the root must then be negative and, indeed, it always is --- for all charged elementary particles, $q/m \gg 1$. Also, an arbitrary accumulation of elementary particles of like charge, trapped by the Reissner--Nordstr\"om field, necessarily leads to $\vert Q \vert > M$. A daemon is, therefore, a naked singularity. \\
We now turn to the study of quantum tunneling of trapped particles through the classically forbidden region between the two turning radii $r_\pm$. \\
The Schr\"odinger equation of one-dimensional motion along the $r$-axis in potential (\ref{pot}) is:
\b
\label{sch}
\frac{d^2 \psi}{dr^2} + \bigl(\frac{A}{r^2} - \frac{B}{r}\bigr)\psi = -\frac{2mE}{\hbar^2} \psi(r) \, ,
\e
where $A = (2m/\hbar^2)a = - mQ^2( 1 - q^2/m^2)/\hbar^2 = $ const $ >0$ and $B = (2m/\hbar^2) b = - 2 m M [1 - (qQ\epsilon)/(mM)]/\hbar^2 > 0$. \\
We are not considering the radial part of a three-dimensional Schr\"odinger equation as we no longer have three-dimensional motion, but an effective one-dimensional problem (the difference between the two in our setup is in the parameter $B$, anyway). \\
Essin and Griffiths \cite{essin} study very thoroughly the potential $-1/x^2$ in quantum mechanics
and its pathologies. They analyze the Shr\"odinger equation
\b
\label{grif}
\frac{d^2 \psi}{dx^2} + \frac{\alpha}{x^2} \psi = - \frac{2mE}{\hbar^2} \psi(x).
\e
When the positive constant $\alpha$ is smaller than 1/4, there are no bound states $(E < 0)$ \cite{essin}. (Negative $\alpha$ would turn the potential into a repulsive one.) \\
The procedure applied in \cite{essin, gupta} for $0 < \alpha < 1/4$, is not suitable for equation (\ref{sch}), due to the presence of the $1/r$ term. Near the origin ($r \to 0$), the term proportional to $1/r$ plays little role and asymptotically equation (\ref{sch}) is the same as equation (\ref{grif}). However, in view of the parameters involved, $q/m \gg 1$, thus $A \gg 1/4$, and such situation cannot be achieved. \\
To address the issue of bound states for equation (\ref{sch}) for the case $A > 1/4$, we will follow the steps of \cite{essin}. We introduce $k^2 = \sqrt{-2mE}/\hbar$. Using the Frobenius method, we search for a solution in the form of power series:
\b
\label{frob}
\psi(r) = r^\nu \sum\limits_{i=0}^{\infty} a_i r^i \, ,
\e
where $a_i =$ const (with $a_0 \ne 0$) and $\nu$ is also a constant. \\
Substituting (\ref{frob}) into equation (\ref{sch}), gives:
\b
\label{sums}
\sum\limits_{i=0}^{\infty} \{ a_i [(i + \nu)(i + \nu - 1) + A]r^{i-2} - a_i B r^{i-1} - k^2 a_i r^i \} = 0 \, .
\e
The coefficients in the different monomials in $r$ must therefore vanish. \\
Setting the coefficient of the $1/r^2$ term to zero yields $\nu (\nu - 1) + A = 0$ or
\b
\label{nu}
\nu = \frac{1}{2} \pm \sqrt{\frac{1}{4} - A} \, .
\e
Here $A \gg 1/4$ and thus $\nu$ is not real. \\
Setting the coefficient of the $1/r$ term to zero gives:
\b
a_1 [\nu (\nu + 1) + A] - a_0 B = 0 \quad \mbox{or} \quad a_1 = \frac{B}{2 \nu}a_0
\e
Setting all other coefficients to zero leads to the recursion relation
\b
\label{rec}
a_{n+2} [(\nu + n + 1)(\nu + n + 2) + A] - a_{n+1} B - k^2 a_n = 0
\e
from which $a_n$ can be determined in terms of $a_0$ for all $n > 2$ . \\
As in the case of \cite{essin}, near the origin ($r \to 0$), the leading term in the solution is $a_0 r^\nu =
a_0 \sqrt{r} \mbox{ exp}[\pm ig \mbox{ ln }r] \, $, where $g = \sqrt{A-1/4}$ is real. The solution near the origin is real, finite (so that $\psi \to 0$ when $r \to 0$) and square-integrable --- it is the same as the one presented in \cite{essin} for equation (\ref{grif}):
\b
\label{bessel}
\psi_k(r) = k \sqrt{\frac{2 \sinh(\pi g)}{\pi g}} \sqrt{r} \, K_{ig}(kr) \, ,
\e
where $K_{ig}$ is the modified Bessel function of order $ig$. \\
The allowed energies are not quantized. The problem with this solution is that there is {\it no ground state}. This means that the particle will cascade down with the release of an unlimited amount of energy. \\
Regarding scattering states ($E > 0$), the general solution to the Schr\"odinger equation (\ref{sch}) near the
origin is, again, as the one given in \cite{essin} for equation (\ref{grif}):
\b
\label{hankel}
\psi_k(r) = \sqrt{r} \, [F H^{(1)}_{ig}(r)+ G H^{(2)}_{ig}(r)] \, ,
\e
where $k = \sqrt{2mE}/\hbar \, , \,  H^{(1,2)}_{ig}(r)$ are Hankel functions, and $F$ and $G$ are constants. The pathology of this solution is in the fact that the boundary condition at $r = 0$ imposes no constraint on the reflection coefficient and does not determine the amplitude of the outgoing wave \cite{essin}. \\
Essin and Griffiths \cite{essin} propose a renormalization procedure in which the potential $U(x) = - a/x^2$ is replaced by the potential
\b
U_\sigma(x) =
\left\{
\begin{array}{ll}
\infty \, , & \mbox{for $x \le \sigma$,} \cr
-a/x^2 \, , & \mbox{for $x > \sigma$\, .}
\end{array}
\right.
\e
The regularized potential has a non-problematic spectrum of discrete bound states and a continuum of scattering states \cite{essin}. Upon taking the limit $\sigma \to 0^+$, all pathologies of the $-1/x^2$ potential resurface. However, if $g$ also tends to zero, together with $\sigma$, then all excited bound states are "squeezed out" into the continuum of scattering states and one single bound ground state with undetermined energy survives \cite{essin}. As mentioned earlier, in view of the parameters of our daemon model, situation in which $g \to 0$ is not achievable. \\
Many other approaches have been considered for resolving the pathologies of the $-1/x^2$ potential, in particular, for maintaining a well-defined vacuum. We will only mention two --- that of Essin and Griffiths \cite{essin}, in which the non-Hermitian Hamiltonian is made self-adjoint by the restriction of its domain, and the one of de Alfaro {\it et al.} \cite{alf}, in which a different combination of conserved charges was chosen as Hamiltonian. The former introduces a free parameter with the dimension of length, thus breaking the scale invariance. The latter leads to breakdown of time-translational invariance. While a single particle in an $-1/x^2$ potential does indeed exhibit pathological properties, we believe that a proper interpretation can rid us of the pathologies. The situation is very similar to the case of a free particle in quantum mechanics. There is no such thing as a free particle with definite energy \cite{griff} --- the wave function does not represent a physically realizable state as it is not normalizable. The physical interpretation of free particles comes in terms of wave packets, where interference of waves of different particles leads to localization and localization leads, in turn, to normalizability. In the case of the $-1/x^2$ potential, the pathologies can be resolved with a likewise many-particle interpretation --- when packets or, rather, ensembles of particles are trapped by the potential. Consider a particle cascading down towards unlimited negative energies. It releases huge amounts of energy which, through particle interactions, excite the other particles of the ensemble. The excited particles, in turn, stop the cascade of the original particle by exciting it with the deposition of positive energy. In other words, the energy exchange between different particles does not allow any particle to "shoot down" towards hugely negative energies and, in result, the particles "bubble up" at bound states of finite energy. \\
In any case, we are interested not in the bound states, but in the continuum of scattering states ($E >0$). Using the Wentzel-Kramers-Brillouin (WKB) approximation method (see \cite{griff}, for example), we will determine the transmission coefficient for tunneling through the classically forbidden region between the two turning radii $r_\pm$. The picture is very similar to the Gamow theory of alpha-decay (see \cite{griff} again). \\
The Schr\"odinger equation (\ref{sch}) can be re-written as:
\b
\frac{d^2 \psi}{dr^2} = - \frac{p^2}{\hbar^2} \psi(r) \, ,
\e
where $p(r) = \sqrt{2m[E - U(r)]}$ is the classical momentum of a particle with energy $E$ moving in potential $U(r)$ (with $E > U(r)$, so that $p(r)$ is real). For tunneling through a potential barrier (namely, across the classically forbidden region between the two turning radii $r_\pm$), the WKB-approximated wave function is given by:
\b
\psi(r) \simeq \frac{D}{\sqrt{\vert p(r) \vert}}\,\, e^{\pm \frac{i}{\hbar}\int\limits_{r_-}^{r_+} \vert p(r) \vert dr} \, ,
\e
where $D =$ const and $\vert p(r) \vert = \sqrt{2m[U(r) - E]}$. \\
The amplitude of the transmitted wave, relative to the amplitude of the incident wave, is diminished by the factor $e^{2 \gamma}$, where
\b
\label{g}
\gamma = \frac{1}{\hbar} \,\, \int\limits_{r_-}^{r_+} \vert p(r) \vert dr
= \frac{2m}{\hbar} \,\, \int\limits_{r_-}^{r_+} \frac{\sqrt{- E r^2 + b r - a}}{r} \, \, dr \, .
\e
The tunneling probability $P$ is proportional to the Gamow factor $e^{-2 \gamma}$ \cite{griff}. \\
To solve the integral in (\ref{g}), we first change the upper limit of integration from $r_+$ to $r_+ - \tau$, where $\tau$ is a small positive parameter. Next, we introduce a new variable $x$, using one of Euler's substitutions:
\b
\label{rp}
\sqrt{- E r^2 + b r - a} = (r - r_+)x \, .
\e
Therefore, $r = (E r_- + r_+ x^2)/(E + x^2)$. (The introduction of $\tau$ does not allow $r$ to reach $r_+$ where both sides of (\ref{rp}) would vanish and the transformation will not be reversible.) Taking the limit $\tau \to 0^+$, allows us to express the original integral as in integral over the $x$-axis from 0 to $- \infty$. Having in mind that the integrand is an even function of $x$, we get:
\b
\gamma = \frac{\sqrt{8m}}{\hbar} \, \frac{E^2 (r_+ - r_-)^2}{r_+} \, \int\limits_{0}^{\infty} \frac{x^2 dx}{(x^2+E)^2(x^2 + E r_-/r_+)} \, .
\e
The integration gives:
\b
\label{gamma}
\gamma = \frac{\pi}{\hbar} \, \sqrt{2mE} \,\, \Bigl( \frac{r_- + r_+}{2} - \sqrt{r_- r_+} \Bigr) \, .
\e
Substituting the turning radii (\ref{r0}) into this equation gives:
\b
\label{gam}
\gamma = -\frac{\pi}{\hbar} \, \sqrt{m} \, \, \vert Q \vert \,\, \sqrt{\frac{q^2}{m^2} - 1}
\,\, + \,\,
\frac{\pi}{\hbar} \, \sqrt{m} \,\, M \,\,
\frac{\epsilon \, \frac{q}{m} \, \frac{Q}{M} - 1}{\sqrt{\epsilon^2 - 1}}\, .
\e
With the drop of the temperature (that is, when $\epsilon$ starts falling from $\infty$ towards 1), the inner turning radius $r_-$ tends to a finite value, while the outer turning radius $r_+$ tends to infinity. The width of the forbidden classical region, $\delta = r_+ - r_- \, $, also tends to infinity in the limit $\epsilon \to 1$:
\begin{center}
\includegraphics[width=10cm]{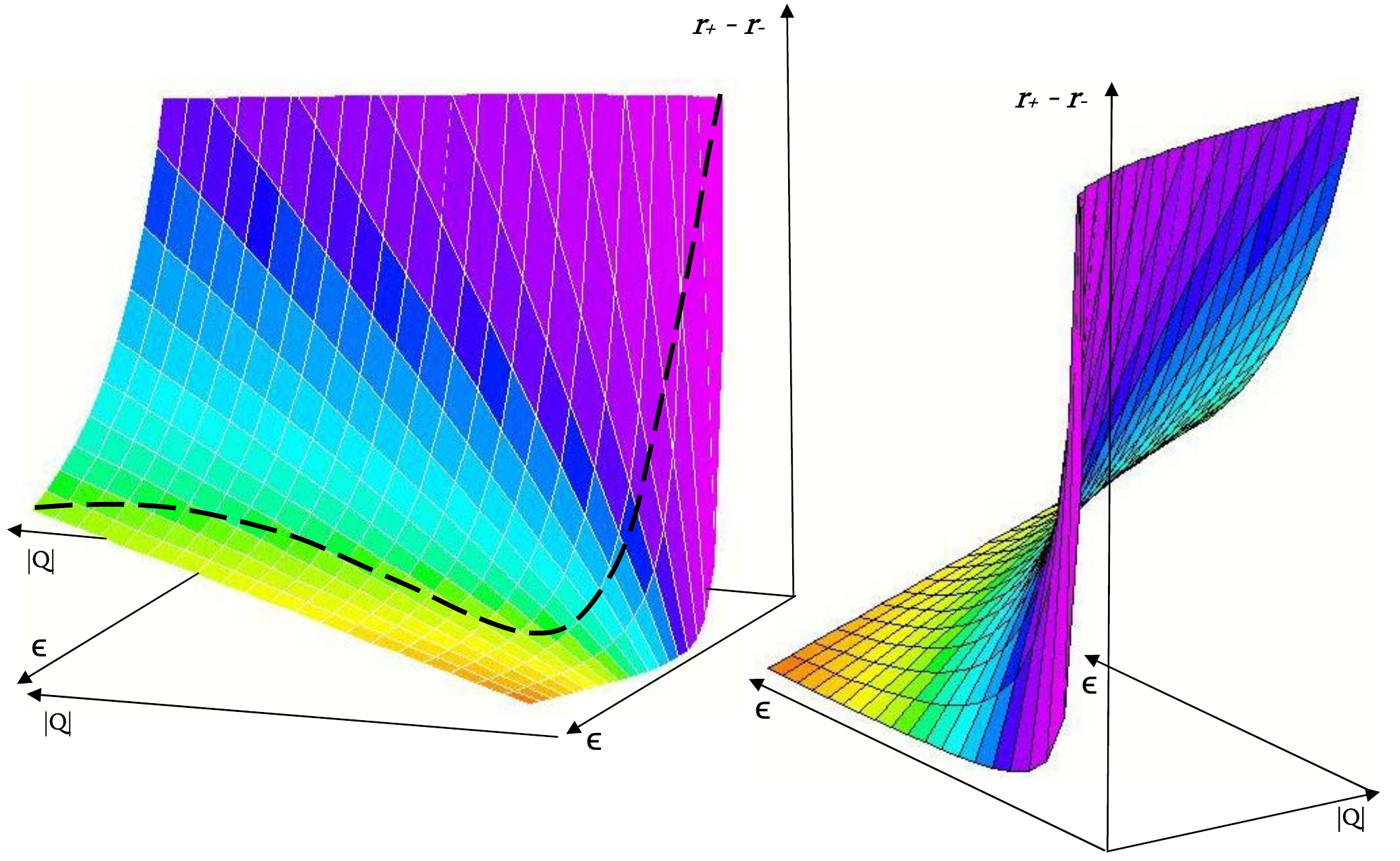}
\vskip.6cm
\noindent
\parbox{130mm}{\footnotesize  {\it Figure 2}: Two perspectives of the three-dimensional plot of width of the forbidden classical region, $\delta = r_+ - r_- \, $, as a function of the daemon charge $\vert Q \vert$ and $\epsilon$ (both $\vert Q \vert$ and $\epsilon$ diminish with time). The dashed curve on the left graph indicates a possible history line obtained by following a path in the $(\vert Q \vert, \epsilon)$-plane. Initially, there is a drop in the width of the barrier followed by a rapid increase to infinity.}
\end{center}
In the very early Universe, at extremely high temperatures (regime $\epsilon \gg 1$), the two turning radii are approximated by:
\b
r_\pm = \frac{qQ \pm m \vert Q \vert}{kT}
\e
and $\gamma$ is not temperature-sensitive:
\b
\label{expa}
\gamma\arrowvert^{\phantom{\frac{\frac{1}{2}}{2}}}_{\epsilon \gg 1^{\phantom{A}}}
= -\frac{\pi}{\hbar} \, \sqrt{m} \, \, \vert Q \vert \,\, \sqrt{\frac{q^2}{m^2} - 1}
\,\, + \,\,
\frac{\pi}{\hbar} \, \frac{qQ}{\sqrt{m}} \, .
\e
As the emitted particles have charge with the same sign as that of the daemon, the absolute value $\vert Q \vert$ of the total charge of the daemon diminishes. The mass $M$ of the daemon  diminishes as well with each emission, but $\vert Q \vert / M \simeq $ const $>1$ at all times. \\
It is, of course, natural to expect that a particle that has just tunneled through the potential barrier of one daemon (and emerged on the "outer" side at $r_+$), would tunnel through into the "inner" side of a neighbouring daemon with the same charge. This happens of course, but not at the rate at which particles tunnel out.  In the case of standard $\alpha$-decay, an $\alpha$-particle is not emitted every time it "knocks" on the "inside wall" of the nucleus. The $\alpha$-particle "rattles" inside the nucleus and the rate of tunneling is given by the Gamow factor, multiplied by the factor $v/2r$, where $v$ is the particle's speed and $r$ is the radius of the nucleus \cite{griff}. The ``bigger'' the nucleus --- the smaller the rate of tunneling. In our model, three objects with the same sign of their charges must be involved in the process of $\alpha$-like emission and re-capture --- two daemons and a particle that tunnels out of one and into the other. The two daemons are very strongly repelled due to the Coulomb interaction between them. In result, the distance between the daemons is much larger than the inner turning radius $r_-$ of each of them. Thus, an emitted particle will have to oscillate between two repelling daemons over distance much greater than the inner turning radius $r_-$ and the rate of re-capture will be much lower than the rate of decay. We will disregard the effect of re-capture of ejectiles on the rate of decrease of $\vert Q \vert$ and $M$.
\begin{center}
\includegraphics[width=12cm]{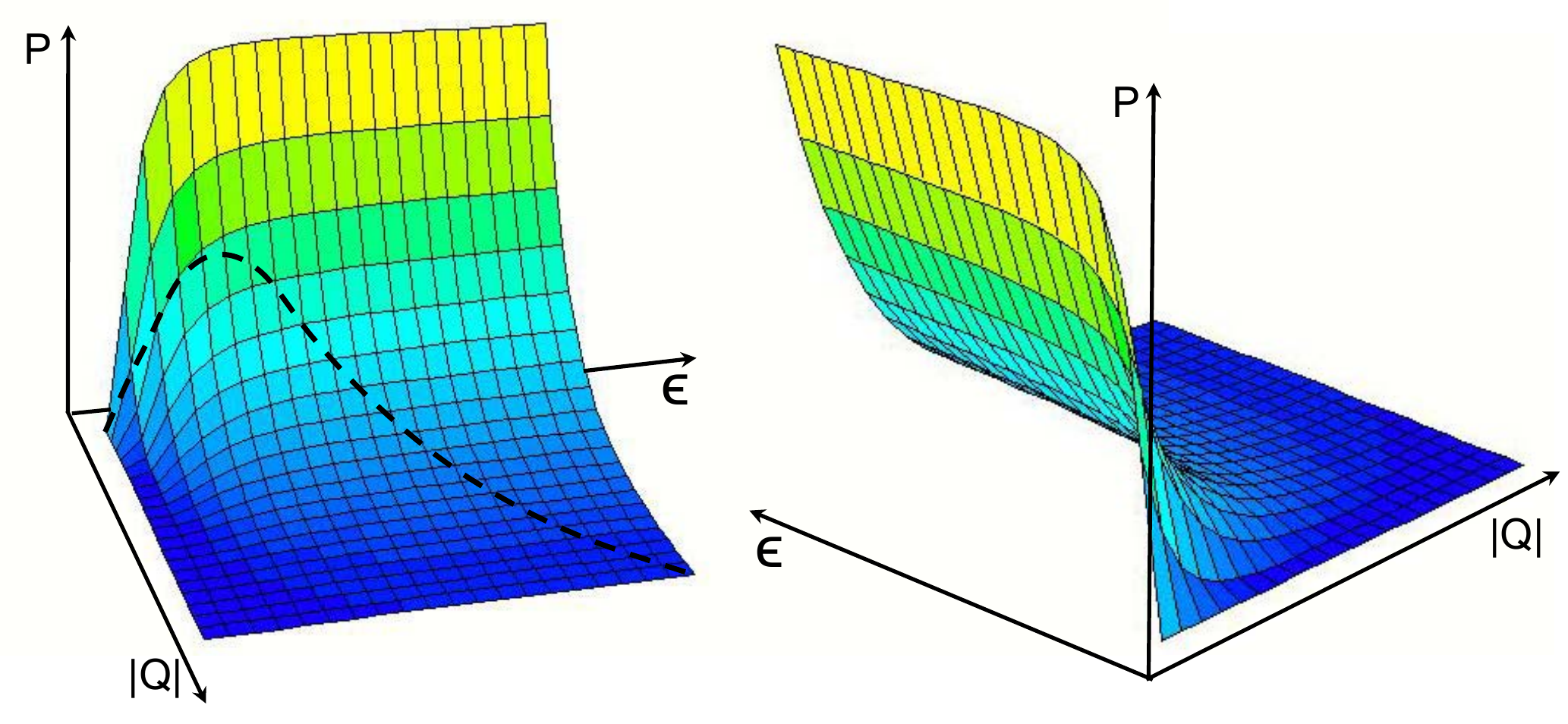}
\vskip.6cm
\noindent
\parbox{130mm}{\footnotesize  {\it Figure 3}: Two perspectives of the three-dimensional plot of the tunneling probability $P$ as a function of the daemon charge $\vert Q \vert$ and $\epsilon$. Again, both $\vert Q \vert$ and $\epsilon$ (or the temperature $T$) diminish with time. The dashed curve on the left graph indicates a possible history line.}
\end{center}
We next expand the first term on the right-hand-side of equation (\ref{expa}) up to first order over the small parameter $m/q$. This gives the probability for tunneling $P$ as proportional to exp$[-(\pi/\hbar) m \sqrt{m} \, \vert Q \vert/ \vert q \vert]$ in the very early Universe (regime $\epsilon \gg 1$) and  growing exponentially with the decrease of $\vert Q \vert$. Over (dimensionless) time $dt$, the charge of the daemon will decrease by the amount  $d \vert Q \vert $ proportional to $-P dt$ and, therefore, in the very early Universe, $\vert Q (t) \vert \simeq \ln (C - t)$,  where $C = \mbox{const}$ and $t$ is dimensionless time. This gives $P(t) \simeq 1/(C - t)$. In alpha-decay, the daughter nucleus recoils after the emission. In view of the analogies between alpha-decay and the current case, we make the following assumption. A particle of kinetic energy $E$ inside the daemon, tunnels through. Tunneling in itself does not change the energy of the particle (otherwise, it would ``resurface'' at point different from the outer turning radius). The recoil energy (needed for conservation of momentum) however, does: the particle's kinetic energy after the emission will be $E$, diminished by the recoil kinetic energy $E_R$. The relativistically correct relation between the linear momentum $p$ of particle of rest mass $m$ and the kinetic energy $E$ of the particle is given by \cite{bethe}: $p^2 = 2 m E + 4 E^2/c^2$ and the recoil kinetic energy $E_R$ is \cite{bethe}: $E_R = (m/M) E + 2(1-m^2/M^2)E^2/M$ where $M$ is the mass of the daughter nucleus. Let us first disregard the relativistic (quadratic in $E$) corrections. Then, if $E_0$ is the total kinetic energy of all particles inside the daemon before the first emission and if we denote $M/m$ by $n$ (figuratively, we have $n$ ``equivalent'' ingredients inside the daemon), then after the first emission, the ejectile will have energy $E_1 = E_0/n - E^{(1)}_R$, where $E^{(1)}_R = E_1 m/(M-m) = E_1/(n-1)$. The energy of the first ejectile is therefore $E_1 = E_0 (n-1)/n^2$. At the same time, as a result of the loss of $m/M$ of the daemon, the total kinetic energy inside the daemon is decreased from $E_0$ to $E_0 - E_0/n = E_0 (n-1)/n$. The second ejectile will have $1/(n-1)$ of this energy, or energy $E_0/n$ prior to leaving the daemon. After tunneling, its energy $E_2$ will be $E_0/n - E^{(2)}_R$, where $E^{(2)}_R = E_2 m/(M-2m) = E_2/(n-2).$ Thus $E_2 = E_0 (n-2)/[n(n-1)]$. The energy inside the daemon is decreased from $E_0 (n-1)/n$ to $E_0 (n-2)/n$. The energy of the third ejectile prior to leaving the daemon will be $1/(n-2)$ of the inside energy, or $E_0/n$. Thus the energy carried away by the third ejectile will be $E_3 = E_0 (n-3)/[n(n-2)]$. The $k^{\mbox{\tiny {\it th}}}$ projectile will therefore have energy $E_k = E_0 (n-k)/[n(n-k+1)] = E_0 (m/M) [1 - 1/(M/m - k + 1)]$. We now take a continuum limit and re-write this as $E(t) = E_0 m/M - E_0/[(m/M)(m/M-k(t)+1)]$, where $k(t)$ is the number of particles emitted after time $t$. The charge inside a daemon decreses in time from its initial value $Q_0$ as $Q(t) = Q_0 - k(t) q$. Thus  $k(t) = Q_0/q - (1/q) \ln (C - t) \simeq M/m - (1/q) \ln (C - t)$. This gives $E(t) \simeq E_0 (m/M) \{1 - 1/[1 + (1/q) \ln (C - t)]\}$. The temperature drops at least as square root of $E(t)$. The outer turning radius $r_+$ (which is inversely proportional to the temperature) has an accelerated increase with time. In result, the scale factor of the Universe, $a(t)$, which is proportional to $r_+$ and, therefore, inversely proportional to the temperature grows with time. The second derivative of $a(t)$ is positive. Therefore we have inflation.
\begin{center}
\includegraphics[width=7cm]{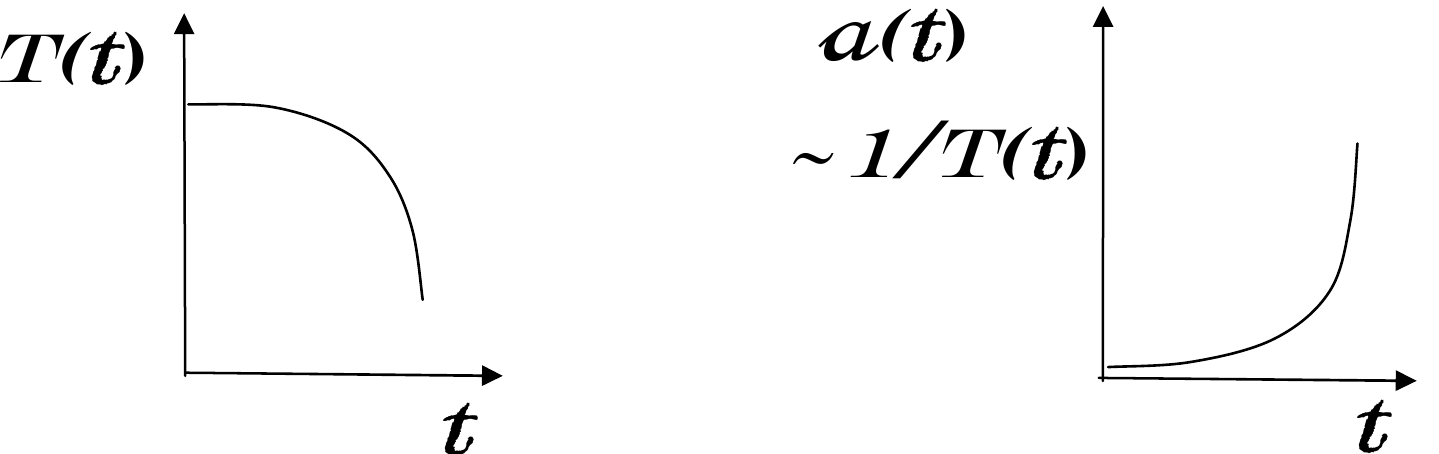}
\vskip.6cm
\noindent
\parbox{120mm}{\footnotesize  {\it Figure 4}: Graph of the temperature $T$ as function of time $t$. The scale factor of the Universe, $a(t)$, is proportional to $r_+$ and inveresly proportional to $T$. The scale factor grows with positive second derivative as $T$ decreases.}
\end{center}
If the relativistic corrections \cite{bethe}, mentioned earlier, were included, than the rapid drop in $T$ would be even more pronounced. \\
Note that in the regime $\epsilon \gg 1$, the width $r_+ - r_- = 2m\vert Q \vert/(kT)$ of the classically forbidden region initially even decreases with time (as the drop of the temperature $T$ is not, initially, as fast as the drop of the charge $\vert Q \vert$ of the daemon --- tunneling is practically temperature-independent). This is when huge amounts of particles gush out of the daemons. The extremely rapid drop in the temperature that follows leads to an extremely rapid growth of $r_+$, together with that of $a(t)$. The ``graceful exit'' of the inflation occurs when the width $r_+ - r_- = 2m\vert Q \vert/(kT)$ of the barrier grows large enough so that quantum tunneling is switched off. This happens before daemons become  fully depleted (bound states inside the daemons should also not be forgotten). In other words, when the temperature $T$ drops sufficiently and the second term in expression (\ref{gam}) for $\gamma$, namely $(\pi/\hbar)M\sqrt{m}[\epsilon q Q/(mM)-1] (\epsilon^2-1)^{-1/2}\, ,$ takes control, a break is put on the tunneling (the lower limit of this term is $(\pi/\hbar) qQ/\sqrt{m}$ when $\epsilon \gg 1$). As the probability for tunneling is brought down very rapidly towards 0 and particles are no longer ejected by the daemons, the medium outside the daemons is no longer cooled by the tunneling process. Without quantum tunneling, the charges of the daemons remain practically constant. However, the temperature of the outside fraction of the Unniverse continues to drop after the rapid accelerated expansion as a different expansion mechanism has naturally taken over. This is the recently proposed Reissner--Nordstr\"om expansion mechanism \cite{pig1, pig2}: with constant charges of daemons, the Universe continues to cool: $T \simeq t^{-1/2}$, and expand: $a \simeq t^{1/2}$. This is the start of the radiation dominated epoch. It is also characterized as the beginning of a supercooling phase. At the end of the inflation, the daemons are still much hotter than the outside fraction of the Universe. A daemon will now cool not through quantum tunneling, but through interaction with the particles of oppositely charged daemons, which, in turn interact with the  particles outside the original daemon. In view of the low densities, this does not happen as fast as the Universe expands. Eventually, the temperature of the daemons and the temperature of the ``free'' fraction of the Universe will equalize and, in result, the Universe will have reheated, but not enough to reignite the inflation (as the daemon temperature now is lower than the one at the end of the inflation and quantum tunneling cannot start). During the reheating, the scale factor $a(t)$ of the Universe does not decrease as there is no mechanism to draw particles, blown away by the growth of the daemons' outer radii, back towards the daemons: the decrease in the outer turning radius $r_+$ of a daemon simply means that particles of the outer fraction will penetrate deeper and deeper into the repulsive field of the daemons. The Universe then enters into another supercooling phase followed by another reheating. This process is repeated until daemons cool down to the temperature of the surrounding fraction and cannot re-ignite futher reheatings. Then the temperature drop will simply follow $T \simeq t^{-1/2}$ and the expansion will be at the rate of $\sqrt{t}$.
\begin{center}
\includegraphics[width=11cm]{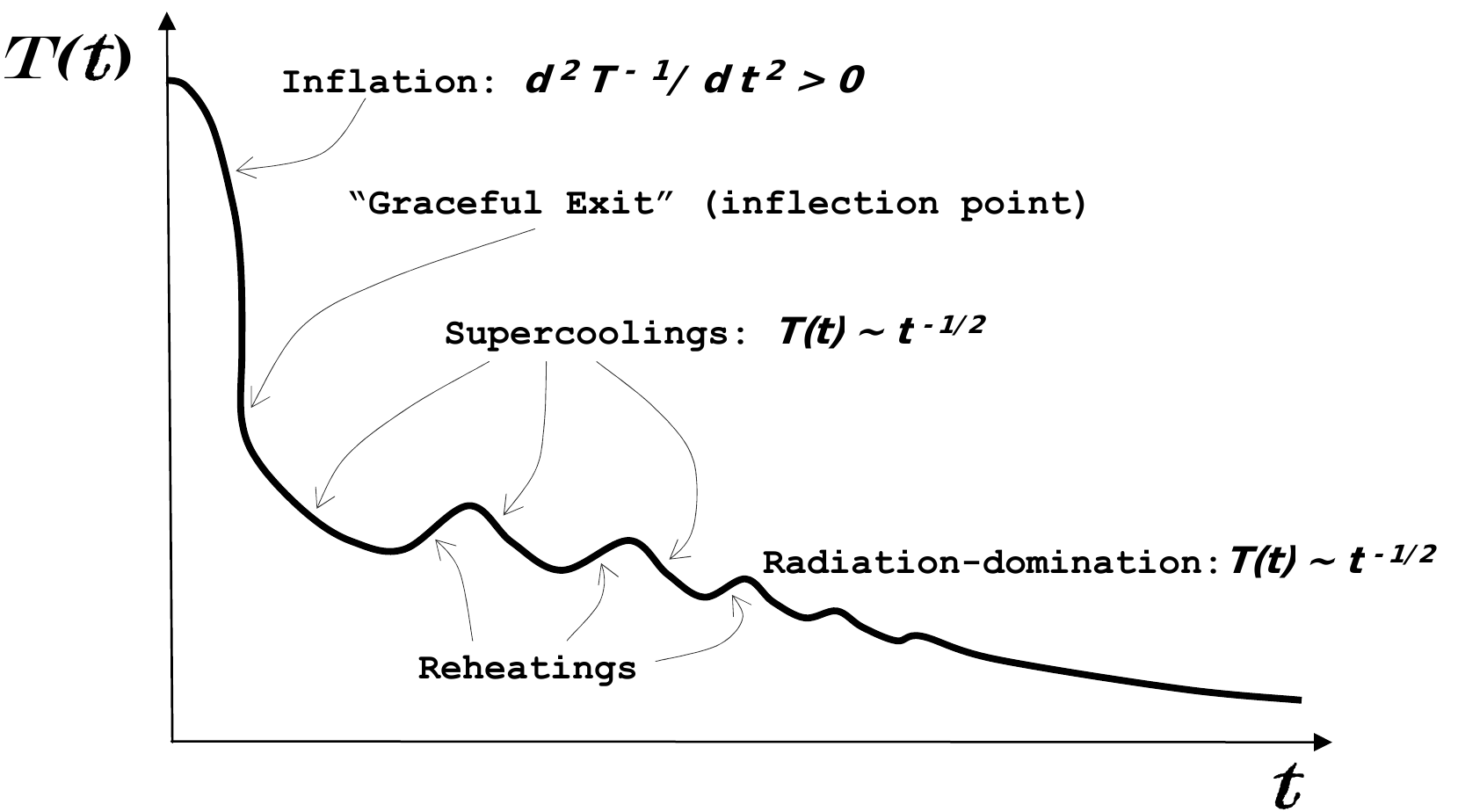}
\vskip.6cm
\noindent
\parbox{130mm}{\footnotesize  {\it Figure 5}: Thermal history of the Universe according to the proposed model. The inflation ($\ddot{a}(t) > 0$) is followed by radiation-dominated epoch characterized by Reissner--Nordstr\"om expansion with series of weaker and weaker reheatings after  supercooling phases. Overall, during radiation domination: $T \simeq t^{-1/2}, \,\,\, a \simeq t^{1/2}$.}
\end{center}
The particles ejected by the daemons, together with all other outside particles, can be viewed as a separate fraction of the Universe,
additional to that of the daemons.  From the viewpoint of an incoming non-daemon particle, such that sign$(Q)q/m \ge -1$, the field of the daemon is characterized by three regions \cite{pig2}. The first one is the attractive region --- from infinity down to the radius $r_c = M (Q^2/M^2 - 1)[1 - (q/m)(Q/M)]^{-1}$, where attraction and repulsion interchange.  There is no gravitationally attractive region for incoming particles such that  $q Q > M m$. The second region is the repulsive region --- between $r_c$ and the outer turning radius $r_+$.  Disregarding tunneling back into daemons (as described earlier), the third region --- between $r_+$ and $r = 0$ --- can be viewed as impenetrable. The Universe can therefore be modelled as a van der Waals gas \cite{pig2} in view of the deep analogies
between the physical picture behind motion in Reissner-Nordstr\"om field and the classical van der Waals molecular model: atoms are
surrounded by imaginary hard spheres and the molecular interaction is strongly repulsive at close proximity, mildly attractive at
intermediate range, and negligible at longer distances. The laws of ideal gas should then be corrected to account for the increased
pressure, due to the additional repulsion, and the decreased available volume, due to the presence of the daemons. \\
We would like to address now the issue of structure formation. This is, essentially,  an initial-data problem. While the proposed inflation model can indeed accommodate a very wide range of initial data, the need of fine tuning with the quantity and composition of matter in the Universe and the nature of primeval inhomogeneities is essential for the study of structure formation. On the other hand, it is important to point out that the study of structure formation for the proposed model is analogous to the study of density perturbations in a van der Waals gas that has undergone accelerated expansion. \\
Replacing the perfect fluid equation of state in cosmology with a van der Waals equation of state has been considered by many authors. The van der Waals quintessence scenario \cite{cap1, cap2} achieves exact accelerated expanding solutions. The van der Waals equation of state allows only observed fluids to be taken into account; phase transitions to occur in the framework of the same evolution; accelerated and decelerated periods to depend on the relative values of the parameters of the state equation with respect to the pressure and matter energy density which are functions of time \cite{cap3}. Additionally, the van der Waals equation of state fits the available astrophysical data with the same accuracy \cite{cap4} as the perfect fluid description which works only for very particular conditions and scales and which is just a rough approximation of cosmic epochs capable of describing stationary situations where phase transitions (which do occur during the evolution of the universe) are not considered. The complicated task of studying structure formation in a van der Waals universe is addressed by Capozziello {\it et al.} in \cite{cap3}. In the redshift range for the van der Waals quintessence model, where presumably structure formation takes place, the baryons energy density dominates over that of the van der Waals fluid that, in this period, is very well approximated by a cosmological constant-like term for all values of the model parameters \cite{cap3}. In the far past, the van der Waals quintessence model, is formally equivalent to the $\Lambda$CDM model with the baryons and the Van der Waals dark matter playing the roles of CDM and $\Lambda$, respectively \cite{cap3}.  This result suggests that structure formation could evolve in a very similar way.
\vskip1cm
\n
It is a pleasure to thank V.G. Gueorguiev and R.I. Ivanov for useful discussions.


\begin{thebibliography}{99}

\bibitem{hawk} S. Hawking, Mon. Not. R. Astr. Soc. {\bf 152}, 75--78 (1971).

\bibitem{rn} H. Reissner, Ann. Phys. ({\it Germany}) {\bf 50}, 106--120 (1916); \newline
G. Nordstr\"om, Proc. Kon. Ned. Akad. Wet. {\bf 20}, 1238--1245 (1918).

\bibitem{mtw} C.W. Misner, K.S. Thorne and J. Wheeler, {\it Gravitation}, W.H. Freeman (1973).

\bibitem{bl} R.H. Boyer and R.W. Lindquist, J. Math. Phys. {\bf 8 (2)}, 265 (1967).

\bibitem{pig1} E.M. Prodanov, R.I. Ivanov, and V.G. Gueorguiev, {\it Reissner--Nordstr\"om Expansion},
Astroparticle Physics {\bf 27} (150--154) 2007, hep-th/0703005.

\bibitem{pig2} E.M. Prodanov, R.I. Ivanov, and V.G. Gueorguiev, {\it Equation of State for a van der
Waals Universe during Reissner--Nordstr\"om Expansion}, Journal of High Energy Physics, JHEP 06(2008)060, arXiv: 0608.0076.

\bibitem{cohen} J.M. Cohen and R. Gautreau, Phys. Rev. {\bf D 19 (8)}, 2273--2279 (1979).

\bibitem{wald} R.M. Wald, {\it General Relativity}, University of Chicago Press (1984).

\bibitem{drob} E.M. Drobyshevski, {\it Detection and Investigation of the Properties of Dark Electric Matter Objects: the First Results and Prospects}, astro-ph/0402367.

\bibitem{essin} A.M. Essin and D.J. Griffiths, {\it Quantum Mechanics of the} $1/x^2$ {\it Potential}, Am. J.
Phys. {\bf 74}(2), 109--117 (2006).

\bibitem{gupta} K.S. Gupta and S.G. Rajeev, {\it Renormalization in Quantum Mechanics}, Phys. Rev {\bf D 48},
5940--5945 (1993).

\bibitem{alf} V. de Alfaro, S. Fubini, and G. Furlan, {\it Conformal Invariance in Quantum Mechanics}, Nuovo Cimento
{\bf 34A}, 569 (1976).

\bibitem{griff} D.J. Griffiths, {\it Introduction to Quantum Mechanics}, Pearson Prentice Hall (2005).

\bibitem{bethe} M. Stanley Livingston and H.A. Bethe, {\it Nuclear Physics --- C. Nuclear Dynamics, Experimental},
Rev. Mod. Phys. {\bf 9(3)}, 245--390 (1937).

\bibitem{cap1} S. Capozziello, S. De Martino, and M. Falanga, {\it Van der Waals Quintessence}, Phys. Lett. {\bf A 299} (5-6), 494--498 (2002).

\bibitem{cap2}  S. Capozziello, S. Carloni, and A. Troisi, {\it Quintessence Without Scalar Fields},  Recent Res. Dev. Astron. Astrophys. {\bf 1}, 625 (2003), astro-ph/0303041.

\bibitem{cap3} S. Capozziello, V.F. Cardone, S. Carloni, S. De Martino, M. Falanga, A. Troisi, and M. Bruni, {\it Constraining Van der Waals Quintessence by Observations}, Phys. Rev. {\bf D 73}, 043512 (2006), astro-ph/0508350.

\bibitem{cap4} V.F. Cardone, C. Tortora, A. Troisi, and S. Capozziello, {\it Beyond the Perfect Fluid Hypothesis for Dark Energy
Equation of State}, Phys. Rev.  {\bf D 73}, 043508 (2006), astro-ph/0511528.

\end{thebibliography}
\end{document}